\documentclass[aps,prd]{revtex4}
\usepackage{amssymb,amsmath,amsfonts,comment,float,latexsym,graphicx,epstopdf}
\usepackage{color}
\usepackage{fancyvrb}
\usepackage{chngcntr}
\usepackage{etoolbox}

\newcommand{\be}{\begin{equation}}
\newcommand{\ee}{\end{equation}}
\newcommand{\ba}{\begin{eqnarray}}
\newcommand{\ea}{\end{eqnarray}}

\newcommand{\grts}{\raise.3ex\hbox{$>$\kern-.75em\lower1ex\hbox{$\sim$}}}
\newcommand{\lets}{\raise.3ex\hbox{$<$\kern-.75em\lower1ex\hbox{$\sim$}}}

\begin{document}

\title{Flavor Changing Leptonic Decays of Heavy Higgs Bosons}

\author{Marc Sher}\email[]{mtsher@wm.edu}
\author{Keith Thrasher}\email[]{rkthrasher@email.wm.edu}

\affiliation{High Energy Theory Group, College of William and Mary, Williamsburg, Virginia 23187, U.S.A.}
\date{\today}

\begin{abstract}

CMS has reported indications (2.4$\sigma$) of the decay of the Higgs boson into $\mu\tau$.    The simplest explanation for such a decay would be a general Two Higgs Doublet Model (2HDM).  In this case, one would expect the heavy neutral Higgs bosons, $H$ and $A$, to also decay in a similar manner.    We study two specific models.   The first is the type III 2HDM, and the second is a 2HDM, originally proposed by Branco et al., in which all flavor-changing neutral processes are given by the weak mixing matrix.  In the latter model, since mixing between the second and third generations in the lepton sector is large, flavor-changing interactions are large.   In this model is found that the decays of $H$ and $A$ to $\mu\tau$ can be as high as 60 percent.
\end{abstract}
\maketitle

\section{Introduction}
Recently, CMS reported\cite{Khachatryan:2015kon} a signal for the Higgs decay $h(125) \to \mu\tau$ at a $2.4\sigma$ level, with a branching ratio of $0.84\pm 0.38$ percent.      Subsequently, ATLAS \cite{Aad:2015gha} reported a signal with a similar central value but larger errors, with a branching ratio of $0.77 \pm 0.62$ percent.    Such a signal, if confirmed in Run 2, would clearly indicate physics beyond the Standard Model (BSM).

Naturally, this has led to a large number of papers explaining the signal in various BSM scenarios.  Some of these include leptoquarks\cite{Cheung:2015yga,Baek:2015mea}, the $331$ model\cite{Hue:2015fbb}, a leptonic dark matter model\cite{Baek:2015fma}, an axion model\cite{Chiang:2015cba}, flavor symmetry models\cite{Campos:2014zaa,Heeck:2014qea} and supersymmetric models\cite{Arganda:2015naa,Arganda:2015uca,Aloni:2015wvn,Zhang:2015csm}.   Some leave the mechanism arbitrary, but explore other ramifications, such as a possible $\bar{t}tH$ excess\cite{Bhattacherjee:2015sia} or an anomaly in $b \to s\mu^+\mu^-$\cite{Crivellin:2015xaa}.

The simplest explanation for a flavor-changing Higgs decay is the general Two Higgs Doublet Model (2HDM) (see Ref. \cite{Branco:2011iw} for an extensive review and list of references).   Several authors have compared, in the context of this model, the expected values for $\tau\to\mu\gamma$, $(g-2)_\mu$ and other lepton number violating processes\cite{Bhattacherjee:2015sia, Davidson:2010xv,Sierra:2014nqa,Dorsner:2015mja,Crivellin:2015hha,He:2015rqa,Altmannshofer:2015esa,Kosnik:2015lka,Omura:2015xcg,Benbrik:2015evd}.   A general 2HDM  has been proposed\cite{Bizot:2015qqo} as an explanation for both $h\to\mu\tau$ and the recent diphoton excess, although this model does have additional fields.

The general 2HDM does have a large number of parameters, and it would be useful to study flavor-changing processes in a more specific context.   In a version of the general 2HDM called Model III, a ansatz motivated by the desire to avoid fine-tuning\cite{Cheng:1987rs} gives flavor-changing couplings in terms of parameters expected to be O(1).   In an even more specific model, by Branco, Grimus and Lavoura (BGL) \cite{Branco:1996bq}, symmetries are used to directly relate the flavor-changing couplings to either the CKM or PMNS matrices, which are measured.   The $h(125)\to\mu\tau$ process was studied in Model III in Ref. \cite{Kosnik:2015lka}, where it was shown that the ansatz does give the correct order of magnitude for the decay.   The process, along with many other flavor-changing processes, in the BGL model was studied in Ref. \cite{Botella:2015hoa}.

2HDMs have two heavy neutral scalars, $H$ and $A$.  If CP is conserved, the $H$ is a scalar and the $A$ is a pseduoscalar.   If the $h \to \mu\tau$ signal is confirmed, then one would expect $H$ and $A$ to also decay into $\mu\tau$.   There are two reasons to expect that the branching ratio of the heavy neutral scalars could be unexpectedly large.   In the alignment (or decoupling) limit of 2HDMs, the gauge boson and fermion couplings of the light Higgs are the same as their SM values.    Thus the mixing parameter $\cos(\alpha-\beta)$ must be small, and yet flavor-changing couplings of the light Higgs will most naturally be proportional to this parameter.   Conversely, flavor-changing couplings of the heavy scalars will be proportional to $\sin(\alpha-\beta)$ and this will not be suppressed.   This fact was pointed out by Altunkaynak, et al.\cite{Altunkaynak:2015twa} in a very detailed analysis of flavor-changing heavy Higgs decays in the hadronic sector.  They briefly mention that $H/A \to \mu\tau$ would be interesting to study since it is unsuppressed by the $\cos(\alpha-\beta)$ factor.   The second reason to expect that the branching ratio might be large is that the flavor-changing interactions in the BGL model will be proportional to the PMNS matrix elements.   Large neutrino oscillations show that 2-3 mixing is maximal, so the 2-3 element of the PMNS matrix is large.  Thus, in the BGL model in particular, one might expect very large rates for $H/A\to\mu\tau$.

	Until very recently, there were no published bounds on $H/A\to \mu\tau$.   A paper by  Buschmann, Kopp, Liu and Wang \cite{Buschmann:2016uzg} appeared in which LHC bounds on $H\to\mu\tau$ from Run 1 are calculated based on the original CMS $h\to\mu\tau$ analysis.  They give results in terms of a generic flavor-changing coupling $\eta_{\mu\tau}$, but don't look at any specific models.    Their work is complementary to ours.  We have not looked at experimental details, but instead will focus on specific models, whereas they do a detailed analysis of the experimental situation.   

	  Shortly after the discovery of the Higgs, Harnik, Kopp and Zupan\cite{Harnik:2012pb} showed that one could extract a bound on $h(125)\to\mu\tau$ from existing bounds on $h(125)\to\tau\tau$.   The bound was $O(10)\%$, but that still gave the better bound  on an $h\mu\tau$ vertex at the time than rare $\tau$ decays.   A similar bound could be derived from $H/A\to\tau\tau$ searches.   While such searches have been carried out, they have all been in the context of a specific supersymmetric model.  In order to have any hope of seeing a signal, it was necessary to enhance the $\tau$ Yukawa coupling with a large $\tan\beta$.  The bounds from CMS\cite{Khachatryan:2014wca} and ATLAS\cite{Aad:2014vgg} typically give an upper bound on $\tan\beta$ of $10-20$ over the mass range for $H$ or $A$ from $150$ GeV to $400$ GeV.    Extraction of a bound for $H/A\to\mu\tau$ would thus be very weak.   This will improve with Run 2 data, but a direct search for $H/A \to\mu\tau$ would be simpler and more reliable.
	
	In the next section, we look at $H/A \to\mu\tau$ in the Type III model, and in the following section study the BGL model.   As noted above, the rate in the latter model can be expected to be large, and we find that to be the case.   Section IV contains our conclusion.

\section{The Type III model}

The requirement that there be no tree-level flavor-changing neutral currents, the Paschos-Glashow-Weinberg theorem\cite{Paschos:1976ay,Glashow:1976nt}, is that all fermions of a given charge must couple to a single Higgs multiplet.    This is generally implemented in a 2HDM by use of a $Z_2$ symmetry.    Without such a symmetry, the Yukawa Lagrangian (involving leptons only) is
\begin{equation}
{\cal L}_Y = - \eta_1\bar{L}_LL_R\Phi_1 - \eta_2\bar{L}_LL_R\Phi_2
\end{equation}
where the $\eta_i$ are real $3\times 3$ matrices.    $\Phi_i$ is given a vacuum expectation value (vev) of $\left({0 \atop v_i}\right)/\sqrt{2}$, and $\tan\beta$ is defined as $v_2/v_1$.    An alternative basis, rotated by an angle $\beta$, has one Higgs, $H_1$ getting a vev and the other $H_2$ not.   In such a basis, $\tan\beta$ doesn't have the usual meaning.   Finally, the third basis is the physical, or mass, basis, in which the scalar mass matrices are diagonalized; this basis is rotated by the angle $\alpha$ relative to the above.    A very detailed description of the various bases was discussed by Davidson and Haber\cite{Davidson:2005cw}.

A nice description of the Yukawa couplings in the type III model was provided by Mahmoudi and Stal\cite{Mahmoudi:2009zx}.   They noted that the above Yukawa Lagrangian gives a mass matrix of
\begin{equation}
M = \frac{v}{\sqrt{2}}(\eta_1 \cos\beta + \eta_2 \sin\beta)
\end{equation}
and then define
\begin{equation}
\kappa\equiv \eta_1 \cos\beta + \eta_2 \sin\beta
\end{equation}
and
\begin{equation}
\rho\equiv -\eta_1 \sin\beta + \eta_2 \cos\beta.
\end{equation}
Thus, $\rho$ does not participate in generating mass for the fermions.  In the Higgs basis, in which only one field gets a nonzero vev, the Lagrangian is
\begin{equation}
{\cal L}_Y = -\kappa \bar{L}_LL_R H_1 - \rho \bar{L}_LL_R H_2
\end{equation}
By construction, $\kappa$ is flavor-diagonal, but the $\rho$ matrix is arbitrary.

Moving to the mass eigenstate basis, they show that the Lagrangian, expanded in terms of neutral fields, becomes
\begin{equation}
-{\cal L}_Y = \frac{1}{\sqrt{2}}\bar{L}\left[ \kappa\  s_{\beta\alpha} + \rho\  c_{\beta\alpha}\right]L h
+\frac{1}{\sqrt{2}}\bar{L}\left[ \kappa\  c_{\beta\alpha} - \rho\  s_{\beta\alpha}\right]L H + \frac{i}{\sqrt{2}}\bar{L}\gamma_5\rho L A
\end{equation}
where $s_{\beta\alpha} = \sin(\beta-\alpha), c_{\beta\alpha} = \cos (\beta-\alpha)$, $h$ is the 125 GeV Higgs, and $H$ and $A$ are the heavy neutral Higgs.  If the couplings of the $h$ are SM-like, then $c_{\beta\alpha}$ must be small.  This Lagrangian shows that the FCNC couplings of the $h$ will be thus suppressed by $c_{\beta\alpha}$, whereas those of the heavy scalars will not be.

The flavor-changing couplings are in the $\rho$ matrix, which, since they have nothing to do with the fermion masses, are arbitrary.    Cheng and Sher\cite{Cheng:1987rs} argued that the most conspicuous feature of the fermion mass matrix is the hierarchical structure, and showed that fine tuning in the Yukawa matrices could be avoided with an ansatz that has become known as the Cheng-Sher ansatz
\begin{equation}
\rho_{ij} = \lambda_{ij}\frac{\sqrt{m_im_j}}{v}
\end{equation}
where the $\lambda_{ij}$ are O(1).   In other words, the flavor-changing couplings are of the order of the geometric mean of the individual Yukawa couplings.   This ansatz has been studied extensively in recent years, and several of the bounds on the $\lambda_{ij}$ are now somewhat less than one.  However, some have argued that the relevant vev is the smaller of the two, leading to a factor of $\tan\beta$ in the effective value of the $\lambda_{ij}$.  Others include an extra factor of $\sqrt{2}$.  In any event, the type III model is generally defined by use of the ansatz, with the $\lambda_{ij}$ of O(1), with the understanding that this is just an order of magnitude estimate.

One can now look at decays of the light Higgs.    The width of the decay into $\bar{\mu}\tau + \bar{\tau}\mu$ is given by
\begin{equation}
\Gamma(h\to\mu\tau) = \lambda_{\mu\tau}^2 c_{\beta\alpha}^2\frac{m_\mu m_\tau m_h}{4\pi v^2}.
\end{equation}
Plugging in the numerical values and dividing by the width of the light Higgs yields
\begin{equation}
B(h\to\mu\tau) = 0.0076 \lambda_{\mu\tau}^2 c_{\beta\alpha}^2
\end{equation}
which is consistent with the CMS central value of $0.0084 \pm 0.0038$ if the product  of $\lambda_{\mu\tau}$ and $c_{\beta\alpha}$ is not too different from 1.   Note that studies of the type I model, for example, allow $c_{\beta\alpha}$ to be as large as $0.4$, so this is not unreasonable.

For the light Higgs decay into $\tau\tau$, one finds
\begin{equation}
B(h\to\tau\tau) = 0.0633 (s_{\beta\alpha} + \lambda_{\tau\tau}c_{\beta\alpha})^2 
\end{equation}
In the alignment limit of $c_{\beta\alpha}=0$, this reduces to the Standard Model result.   Note that there are currently large uncertainties in the $h\to\tau\tau$ experimentally measured branching ratios, and a $20-30\%$ deviation could easily be accommodated as long as $\lambda_{\tau\tau}$ is not too large.  Thus, keeping in mind that the $\lambda_{ij}$ are order of magnitude, one sees that this model can account for the observed results in light Higgs decays.

But we are interested in heavy Higgs decays, and ratios of branching ratios can be calculated.   For the moment, consider the alignment limit (the results will then apply to the pseudoscalar as well).   In this case, one finds
\begin{equation}
\frac{B(H\to\mu\tau)}{B(H\to\tau\tau)} = \frac{m_\mu}{m_\tau} \frac{\lambda_{\mu\tau}^2}{\lambda_{\tau\tau}^2}.
\end{equation}
Since the ratio of $\lambda_{\mu\tau}$ to $\lambda_{\tau\tau}$ must be somewhat larger than one, this is at least $6\%$ and could be substantially higher.   In the alignment limit, there is no coupling to vector bosons, thus the only other substantial decay is $H\to\bar{b}b$, and
\begin{equation}
\frac{B(H\to\tau\tau)}{B(H\to\bar{b}b)} = \frac{m_\tau}{3m_b} \frac{\lambda_{\tau\tau}^2}{\lambda_{bb}^2}.
\end{equation}
If the $\lambda$'s are equal, this will be the same as the ratio of branching ratios for the light Higgs, or approximately $11\%$, although this number will have large uncertainties.   This will not be qualitatively changed by moving away from the alignment limit.
For the heavy Higgs in the model, we thus see that it is unlikely that the $\mu\tau$ decay mode will dominate,   However, it will likely be substantially higher than the $0.8\%$ branching ratio for the light Higgs.

It was noted earlier that very recent results from Buschmann, Kopp, Liu and Wang \cite{Buschmann:2016uzg} are complementary to ours in that they look at experimental bounds.   They give bounds from the 8 TeV LHC run on a possible flavor-changing coupling.  and consider both LHC constraints from $H/A$ decays as well as constraints from $\tau\to\mu\gamma$.    In our notation, they show that the preferred values of $\rho_{\mu\tau}$  are between $0.004$ and $0.02$.   From Equation 7, this gives a value of $\lambda$ between $2$ and $12$.   However, their technique will be very valuable in LHC Run 2, where much tighter bounds can be obtained.

The BGL model is a very different model with much less uncertainty in the results, since the mixing is directly related to the PMNS matrix.   We now turn to that model.

\section{The BGL Model}
In a general 2HDM the Yukawa Lagrangian involving only quark fields  takes the form
\be
{\mathcal{L}}_{Y} =
-\overline{Q_{L}^{0}}\,\big[\Gamma_{1}\,\Phi_{1}+\Gamma_{2}\,\Phi_{2}\big]\,d_{R}^{0}
-\overline{Q_{L}^{0}}\,\big[\Delta_{1}\,\tilde{\Phi}_{1}+\Delta_{2}\,\tilde{\Phi}_{2}\big]\,u_{R}^{0} +H.c.,
\ee
where $\Gamma_i$ and $\Delta_i$ are the Yukawa coupling of the quarks. BGL showed\cite{Branco:1996bq} by imposing a discrete symmetry on the fields,
\be
Q_{Lk}^{0}\mapsto \exp {(i\tau)}\, Q_{Lk}^{0}\, ,\qquad
u_{Rk}^{0}\mapsto \exp {(i2\tau)}\,u_{Rk}^{0}\, ,\qquad \Phi
_{2}\mapsto \exp {(i\tau)}\,\Phi_{2}\, ,  \label{usym}
\ee
where $\tau \neq 0, \pi$, with all other quark fields transforming 
trivially under the symmetry, one could have the Yukawa interactions completely determined by the CKM matrix V. The index $j$ can be fixed as either 1,2 or 3. An alternative symmetry can be chosen where the fields transform as
\be
Q_{Lk}^{0}\mapsto \exp {(i\tau)}\, Q_{Lk}^{0}\, ,\qquad
d_{Rk}^{0}\mapsto \exp {(i2\tau)}\, d_{Rk}^{0}\, ,\quad 
\Phi_{2}\mapsto \exp {(-i\tau)}\, \Phi_{2}\, .  \label{dsym}
\ee
The set of symmetry transformations given in Eq. (\ref{usym}) leads to FCNC contained only in the down sector, while the transformation in Eq.(\ref{dsym}) gives rise to FCNC in the up sector.   This leads, depending on the value of $k$, to six possible models.  Similarly, one can have the same possibilities applied in the lepton sector, leading to FCNC in the charged lepton sector.    These models are referred to as $\nu_j$ models.

The Yukawa couplings of the light Higgs can be derived following Refs. \cite{Branco:1996bq} and \cite{Botella:2015hoa}.    Their result for the Yukawa coupling to $\mu\tau$ is
\be
Y_{\mu\tau}= -U_{\mu j}^*U_{\tau j} \frac{M_\tau}{v}c_{\beta\alpha}(t_\beta + t_\beta^{-1})
\ee
where there is no sum on $j$ and the values of $j=1,2,3$ correspond to three possible models.
Here we see the attractive feature of BGL models.   The flavor-changing couplings are given by the elements of the PMNS matrix, and thus are determined only by the usual mixing angles in 2HDMs.

The decay width of $h \rightarrow \bar{\mu}\tau+\bar{\tau}\mu$ in the $\nu_j$ type model is then,
\be
\Gamma(h\to\mu\tau) =\Gamma_{sm}\left(h\rightarrow \bar{\tau}\tau \right) c^2_{\beta\alpha} 
\left (t_\beta+t^{-1}_\beta \right)^2 \left|U_{\tau j} U_{\mu j} \right|^2
\ee
where $\Gamma_{sm}\left(h\rightarrow \bar{\tau}\tau \right)=\frac{m^2_\tau m_h}{8 \pi v^2}$. 

From the measured decay width (using CMS results) one can now plot the allowed region in the $t_\beta - c_{\beta\alpha}$ plane.    This is done in the left figure of Figure 1, with one and two standard deviation bands plotted.    Note that the alignment limit of $c_{\beta\alpha}=0$ is excluded since the CMS branching ratio is more than $2\sigma$ away from zero.

Of course, the LHC data from Run 1 does not allow the properties of the Standard Model Higgs to deviate too much from the alignment limit.   There have been many studies of the allowed range in 2HDM models (see Ref. \cite{Craig:2013hca} for an extensive list of references).     Since the quark and gauge boson sectors of this model are very similar to the Type I 2HDM, the parameter-space can be restricted by this data.    In right side of Figure 1, we have shown the region allowed by the LHC Run1 data in the Type 1 model.     This will be slightly modified in the BGL model.   The couplings of the vector bosons in the Type 1 and BGL models are the same.   The coupling to quarks in the BGL model is the same as in the Type 1 model times $\sin^2\beta + \sin\beta\cos^2\beta$.    For $\tan\beta > 2$, this gives a discrepancy of a few percent, which is negligible.   As a result, the full analysis in the BGL model will be virtually indistinguishable from the bounds on the right side of Figure 1.  To a good approximation, for most values of $\tan\beta$, one requires (at $2\sigma$) only that $|\cos_{\beta\alpha}| <  0.4$ and we will thus restrict our discussion to those values.

\begin{figure}[h]
\centering
\includegraphics[width=8cm, height=5cm]{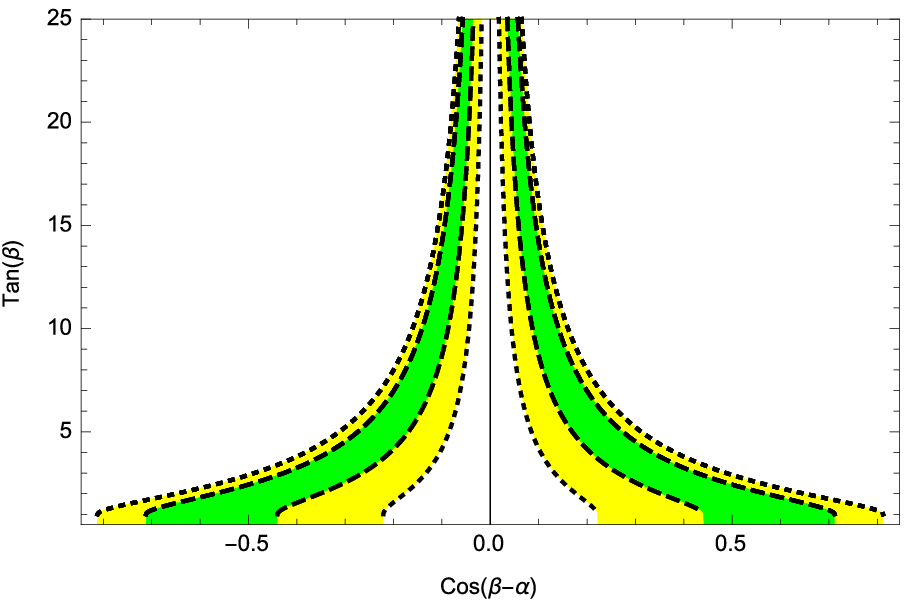}
\hspace{8mm}
\includegraphics[width=8cm, height=5cm]{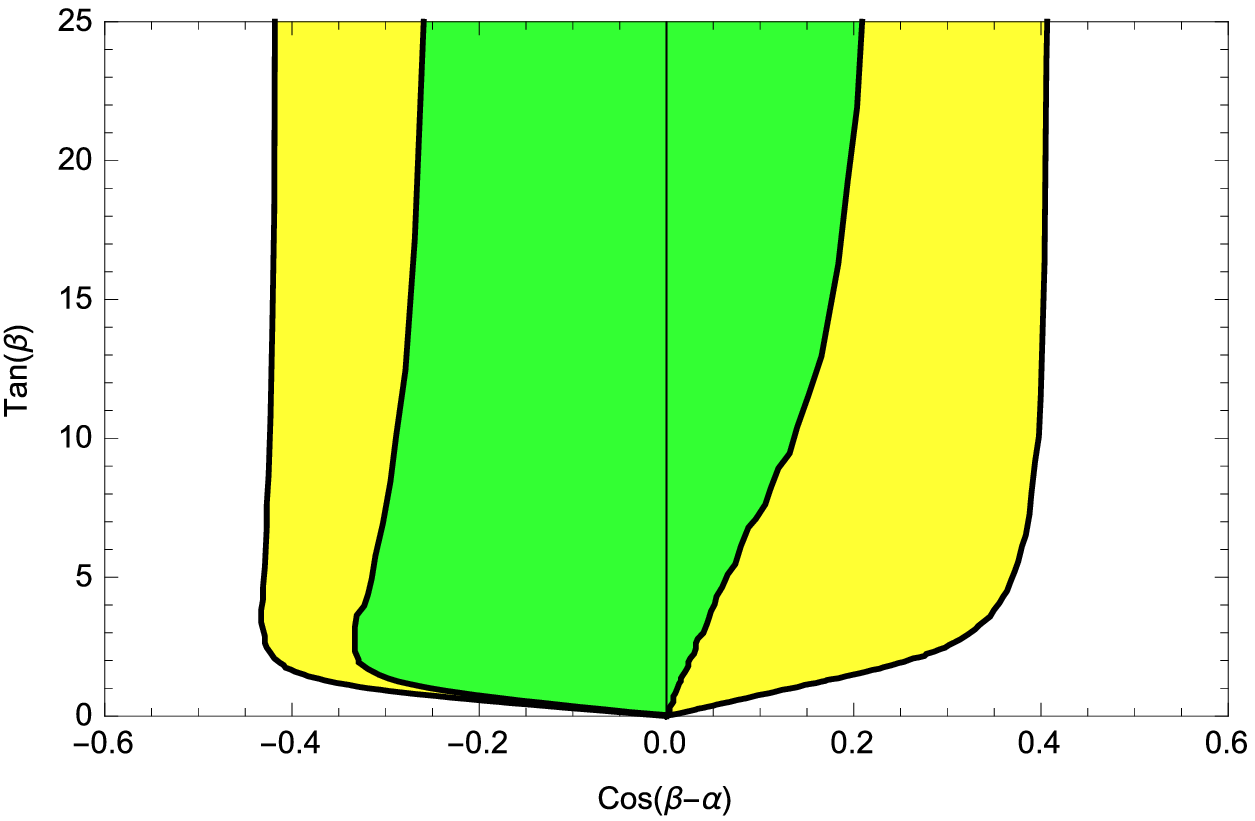}
\caption{(Left) Plot of allowed region for $\tan\beta$ as a function of $\cos(\beta -  \alpha)$ for  $h\to\mu\tau$ in the $(\nu_3,t)$-type BGL model using $1\sigma$ and $2\sigma$ confidence intervals. (Right) Bounds placed on $\tan\beta$ and $\cos(\beta - \alpha)$ for the Type-I 2HDM using data from LHC Run1.}
\end{figure}

We now turn to the couplings of the heavy Higgs.    It is straightforward to calculate the width of the heavy Higgs bosons in the model.  We are choosing a value for the heavy Higgs mass of $350$ GeV.  If it is heavier, the decay into top quark pairs will dominate the decays, leading to very small branching ratios.   Below $350$ GeV, the masses cancel in branching ratios, except for phase space in decays to pairs of gauge bosons.   However, these decays are suppressed by $c_{\beta\alpha}^2$ for $H$ and vanish for $A$, and thus the results are not very sensitive to the mass chosen.  The results are in Table 1.    Not surprisingly, the flavor-changing decays are proportional to the leptonic mixing angles and $s_{\beta\alpha}^2$, which are not small.     From these widths, one can calculate the branching ratio of $H/A\to\mu\tau$.     Note that the branching ratio of the $A$ is independent of $c_{\beta\alpha}$.   The results in Table 1 depend on the model chosen - one can set $j=1,2,3$ and $k=1,2,3$ independently.  Note that for $k=1,2$, the b-quark coupling scales as $\tan\beta$ (instead of $1/\tan\beta$ for $k=3$).  Thus, the b-quark coupling will not be suppressed, and the branching ratio to $\mu\tau$ for either $H$ or $A$ will be very similar to that of Model III in the last section.  It will never be particularly large.  We will thus focus on the $k=3$ models.

The most interesting cases are when $j=2,3$.   For $j=1$, the PMNS mixing angles are smaller.   The value of $|U_{\tau j}|$ and $|U_{\mu j}|$ are between 0.45 and 0.77 for $j=2,3$.   Since $V_{tb}$ is very close to one, the b-quark coupling is very small for large $\tan\beta$.    Thus, for example, the width for $A\to\bar{b}b$ becomes small for large $\tan\beta$ (in the $k=3$ model), leading to very large branching ratios for $A\to\mu\tau$.    We are not including a possible decay of the $H$ into two Higgs bosons since it depends on unknown scalar self-couplings (there is no such coupling for the $A$).  

In the left side of Figure 2, we plot the branching ratio for $H\to\mu\tau$ and $A\to\mu\tau$ in the $j=k=3$ model.  The solid (dashed) lines correspond to $H$ ($A$) decays.    One can see that huge branching ratios for $H\to\mu\tau$ will occur for a large part of the allowed parameter-space, and for virtually all of the parameter-space, the branching ratio for $A\to\mu\tau$ will be very large.   In the right side of Figure 2, we plot the same for $j=2, k=3$.  Here the branching ratios are a little smaller because the $(3,2)$ element of the PMNS matrix is smaller than the $(3,3)$ element.    

Thus, in one version of the BGL model, the branching ratios to $\mu\tau$ in the allowed parameter space can be quite large, over $60\%$.    This will certainly have a substantial impact on the experimental searches for these states.

\begin{table}[H]
\centering
\begin{tabular}{|c|c|c|}
\hline 
X & $H$ & $A$\tabularnewline
\hline 
\hline 
$\Gamma(X\rightarrow\mu\tau)$ & $\frac{m^2_\tau m_H}{8 \pi v^2}s_{\beta\alpha}^{2}\left(t_{\beta}+t_{\beta}^{-1}\right)^{2}\left|U_{\tau j}U_{\mu j}\right|^{2}$ & $\frac{m^2_\tau m_A}{8\pi v^2} \left(t_{\beta}+t_{\beta}^{-1}\right)^{2}\left|U_{\tau j}U_{\mu j}\right|^{2}$\tabularnewline
\hline 
$\Gamma(X\rightarrow\bar{\tau}\tau)$ & $\frac{m^2_\tau m_H}{8 \pi v^2}\left(c_{\beta\alpha}-s_{\beta\alpha}\left(t_{\beta}-\left(t_{\beta}+t_{\beta}^{-1}\right)\left|U_{\tau j}\right|^{2}\right)\right)^{2}$ & $\frac{m^2_\tau m_A}{8 \pi v^2}\left(t_{\beta}-\left(t_{\beta}+t_{\beta}^{-1}\right)\left|U_{\tau j}\right|^{2}\right)^{2}$\tabularnewline
\hline 
$\Gamma(X\rightarrow\bar{b}b)$ & $\frac{3 m^2_b m_H}{8 \pi v^2}\left[c_{\beta\alpha}-s_{\beta\alpha}\left(t_{\beta}-\left(t_{\beta}+t_{\beta}^{-1}\right)\left|V_{kb}\right|^{2}\right)\right]^{2}$ & $\frac{3 m^2_b m_A}{8 \pi v^2}\left[t_{\beta}-\left(t_{\beta}+t_{\beta}^{-1}\right)\left|V_{kb}\right|^{2}\right]^{2}$\tabularnewline
\hline 
$\Gamma(X\rightarrow W^{\pm}W^{\mp})$ & $\frac{m_{H}^{3}c_{\beta\alpha}^{2}}{16\pi v^{2}}\left(1-4\left(\frac{m_{W}}{m_{H}}\right)^{2}+12\left(\frac{m_{W}}{m_{H}}\right)^{4}\right)\sqrt{1-4\left(\frac{m_{W}}{m_{H}}\right)^{2}}$ & 0 \tabularnewline
\hline 
$\Gamma(X\rightarrow ZZ)$ & $\frac{m_{H}^{3}c_{\beta\alpha}^{2}}{32\pi v^{2}}\left(1-4\left(\frac{m_{W}}{m_{Z}}\right)^{2}+12\left(\frac{m_{W}}{m_{Z}}\right)^{4}\right)\sqrt{1-4\left(\frac{m_{W}}{m_{Z}}\right)^{2}}$ & 0\tabularnewline
\hline 
\end{tabular}
\caption{Decay widths of for the heavy scalar Higgs bosons in the $(\nu_j,u_k)$-type BGL models.} 
\end{table}

\begin{figure}[h]
\centering
\includegraphics[width=8.5cm, height=6cm]{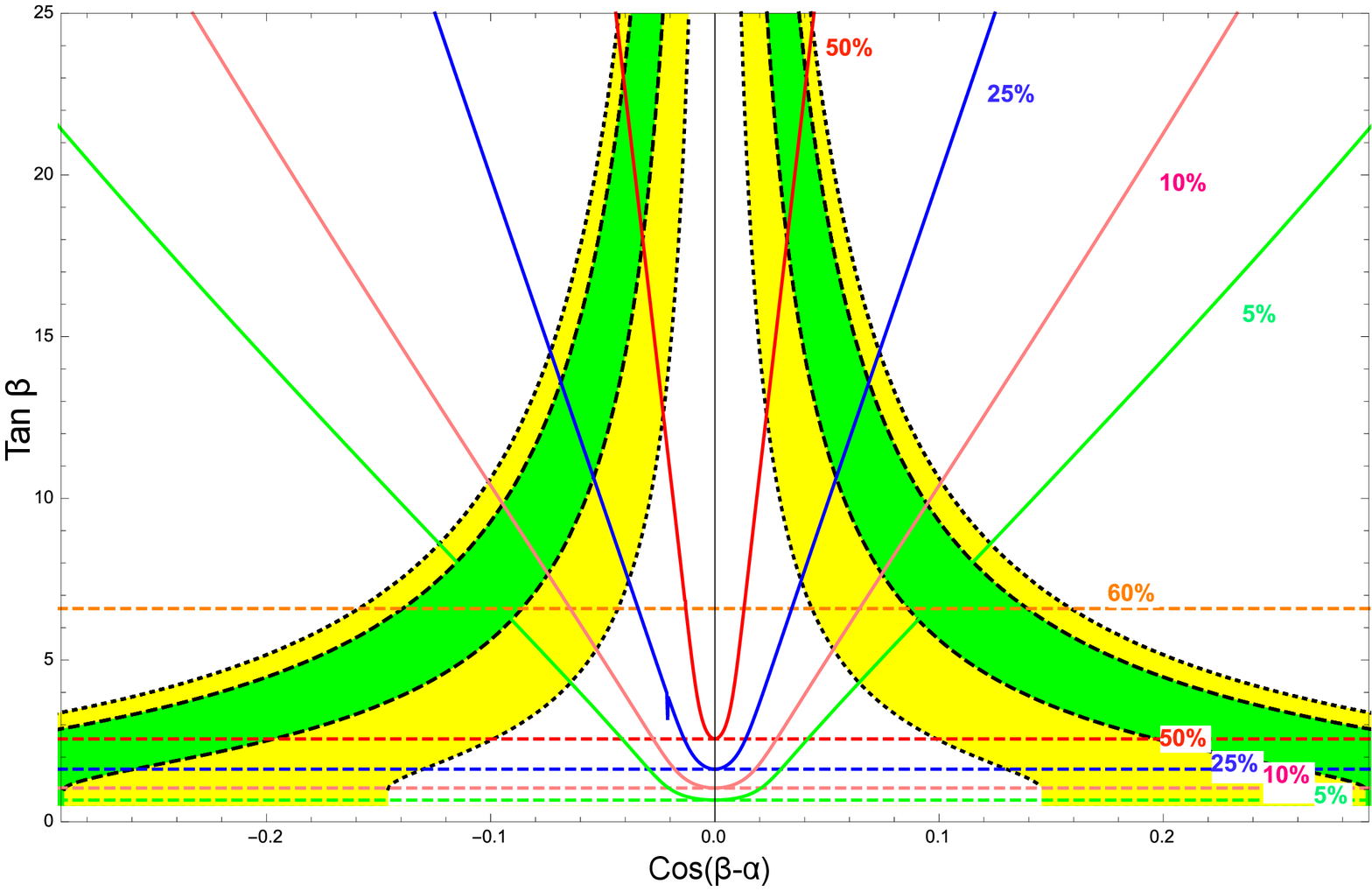} 
\includegraphics[width=8.5cm, height=6cm]{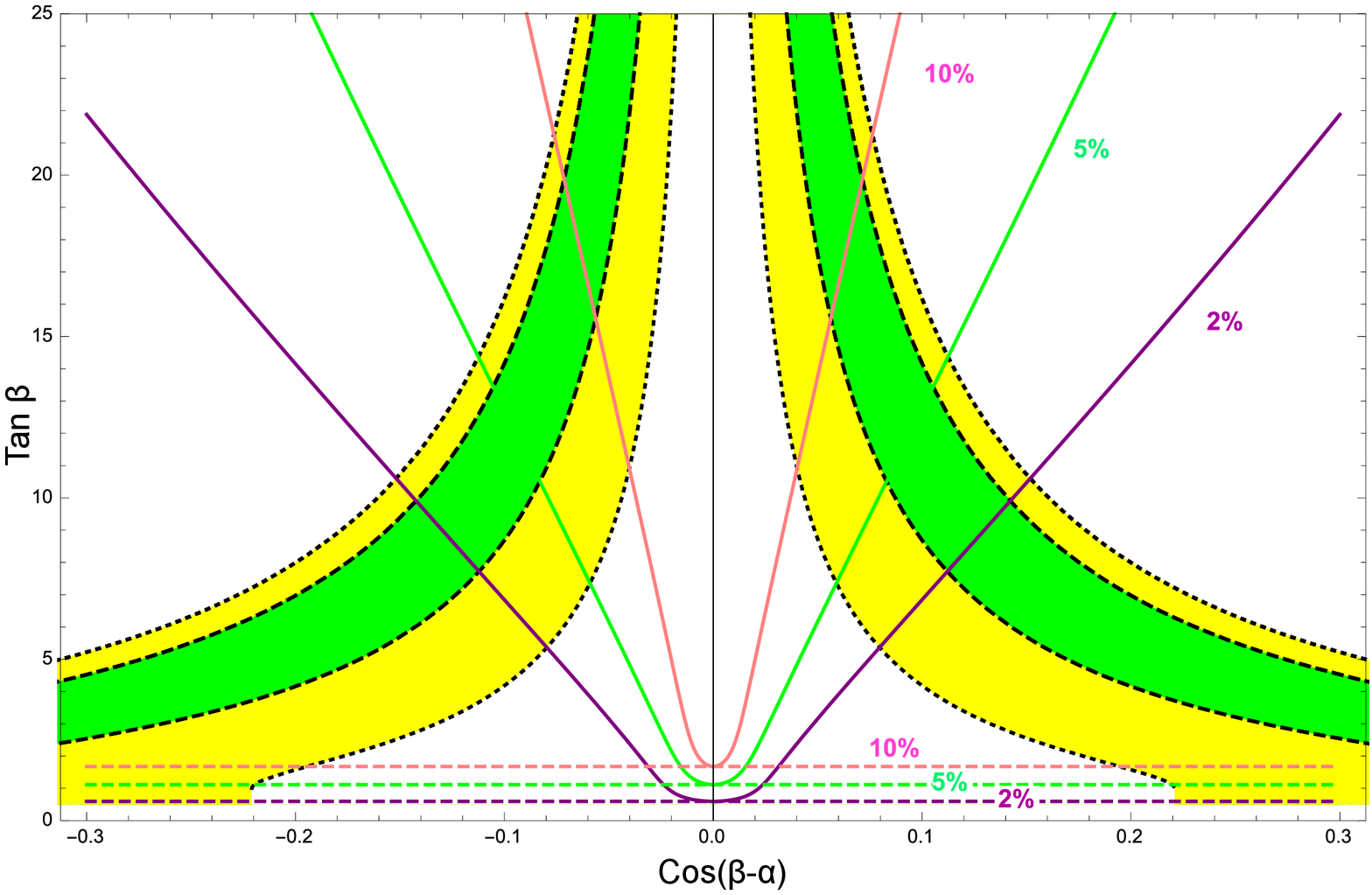} 
\caption{Composite of plots of the bounds for $h/H/A\to\mu\tau$.  The left plot shows bounds on $\tan\beta$ and $\cos (\beta-\alpha)$ in the $(\nu_3,t)$-type BGL model. The right plot show bounds in the $(\nu_2,t)$-type BGL model. Green and yellow bands are bounds at $1\sigma$ and $2\sigma$ level from $h\to\mu\tau$ using CMS data. Solid (Dashed) lines are contours for $H\to\mu\tau$ ($A\to\mu\tau$) at the various branching fractions labeled in the plots. In each case the Higgs masses $m_A$ and $m_H$ were chosen to be $350$ GeV.}
\end{figure}

\section{Results}

Should the CMS indications for a nonzero branching ratio for $h\to\mu\tau$ be confirmed in Run 2, the most likely culprit will be a Two-Higgs Doublet Model.    This would imply a nonzero branching ratio for the heavy neutral scalars in the model.    The recent analysis of Buschmann, et al. \cite{Buschmann:2016uzg} shows that one can extract some bounds on $H\to\mu\tau$ from the CMS search, but a dedicated search for the decay mode in Run 2 could be quite valuable.      In general, the flavor-changing neutral couplings can be arbitrary, but can be tightly constrained in particular models, although one would expect the suppression by $c_{\beta\alpha}$ in $h$ decay to be absent in $H$ and $A$ decays.  

We have examined two such models, Model III and the BGL model.  Are there any other models that might have a large rate?   In the conventional 2HDMs, there are no tree level FCNC and thus flavor-changing decays can only occur through a loop.   This will cause a substantial suppression in the branching ratios.    It has been noted that supersymmetric models with large smuon-stau mixing can at large $\tan\beta$  yield a relatively large rate for \cite{Babu:2002et}  $\tau\to 3\mu$ and for \cite{Sher:2002ew} $\tau\to\eta\mu$, due to a $\tan^6\beta$ dependence but the rates are still quite small and also go through a loop.   We know of no other models which are predictive and can yield a large branching fraction for $H/A \to\mu\tau$.

 In Model III, the ratio of $H\to\mu\tau$ to $H\to\tau\tau$ will be at least $6\%$ and could be much higher, and the latter will have a branching ratio of roughly $10\%$.    In the BGL model, there is an additional enhancement since the flavor-changing couplings are proportional to the PMNS matrix, which has very large mixing in the 2-3 sector.   We have seen that branching ratios for $H\to\mu\tau$ and $A\to\mu\tau$ can be as large as $60\%$.

 \vskip 1.0cm
{\parindent = 0pt {\bf Acknowledgments}}
 \vskip 0.5cm
The authors would like to thank Chris Carone, Josh Erlich, Margarida Rebelo and Rui Santos for useful discussions.   This work was supported by the NSF under Grant PHY-1519644.


\newpage


    \begin{thebibliography}{99}
    
\bibitem{Khachatryan:2015kon} 
  V.~Khachatryan {\it et al.} [CMS Collaboration],
  Phys.\ Lett.\ B {\bf 749}, 337 (2015)
  doi:10.1016/j.physletb.2015.07.053
  [arXiv:1502.07400 [hep-ex]].
\bibitem{Aad:2015gha} 
  G.~Aad {\it et al.} [ATLAS Collaboration],
  JHEP {\bf 1511}, 211 (2015)
  doi:10.1007/JHEP11(2015)211
  [arXiv:1508.03372 [hep-ex]].
\bibitem{Cheung:2015yga} 
  K.~Cheung, W.~Y.~Keung and P.~Y.~Tseng,
  arXiv:1508.01897 [hep-ph].
\bibitem{Baek:2015mea} 
  S.~Baek and K.~Nishiwaki,
  arXiv:1509.07410 [hep-ph].
  E as of 25 Dec 2015
\bibitem{Hue:2015fbb} 
  L.~T.~Hue, H.~N.~Long, T.~T.~Thuc and N.~T.~Phong,
  arXiv:1512.03266 [hep-ph].
\bibitem{Baek:2015fma} 
  S.~Baek and Z.~F.~Kang,
  arXiv:1510.00100 [hep-ph].
\bibitem{Campos:2014zaa} 
  M.~D.~Campos, A.~E.~C.~Hernández, H.~Päs and E.~Schumacher,
  Phys.\ Rev.\ D {\bf 91}, no. 11, 116011 (2015)
  doi:10.1103/PhysRevD.91.116011
  [arXiv:1408.1652 [hep-ph]].
\bibitem{Heeck:2014qea} 
  J.~Heeck, M.~Holthausen, W.~Rodejohann and Y.~Shimizu,
  Nucl.\ Phys.\ B {\bf 896}, 281 (2015)
  doi:10.1016/j.nuclphysb.2015.04.025
  [arXiv:1412.3671 [hep-ph]].
\bibitem{Chiang:2015cba} 
  C.~W.~Chiang, H.~Fukuda, M.~Takeuchi and T.~T.~Yanagida,
  JHEP {\bf 1511}, 057 (2015)
  doi:10.1007/JHEP11(2015)057
  [arXiv:1507.04354 [hep-ph]].
\bibitem{Arganda:2015naa} 
  E.~Arganda, M.~J.~Herrero, X.~Marcano and C.~Weiland,
  arXiv:1508.04623 [hep-ph].
\bibitem{Arganda:2015uca} 
  E.~Arganda, M.~J.~Herrero, R.~Morales and A.~Szynkman,
  arXiv:1510.04685 [hep-ph].
\bibitem{Aloni:2015wvn} 
  D.~Aloni, Y.~Nir and E.~Stamou,
  arXiv:1511.00979 [hep-ph].
\bibitem{Zhang:2015csm} 
  H.~B.~Zhang, T.~F.~Feng, S.~M.~Zhao, Y.~L.~Yan and F.~Sun,
  arXiv:1511.08979 [hep-ph].
\bibitem{Bhattacherjee:2015sia} 
  B.~Bhattacherjee, S.~Chakraborty and S.~Mukherjee,
  arXiv:1505.02688 [hep-ph].
\bibitem{Crivellin:2015xaa} 
  A.~Crivellin, G.~D'Ambrosio and J.~Heeck,
  arXiv:1505.02026 [hep-ph].

\bibitem{Branco:2011iw} 
  G.~C.~Branco, P.~M.~Ferreira, L.~Lavoura, M.~N.~Rebelo, M.~Sher and J.~P.~Silva,
  Phys.\ Rept.\  {\bf 516}, 1 (2012)
  doi:10.1016/j.physrep.2012.02.002
  [arXiv:1106.0034 [hep-ph]].
  \bibitem{Davidson:2010xv}
  S. ~Davidson and G.-J.~Grenier
  Phys.\ Rev.\ D {\bf 81}, 095016 (2010)
  doi:10.1103/PhysRevD.81.095016 [arXiv:1001.0434 [hep-ph]]
\bibitem{Sierra:2014nqa} 
  D.~Aristizabal Sierra and A.~Vicente,
  Phys.\ Rev.\ D {\bf 90}, no. 11, 115004 (2014)
  doi:10.1103/PhysRevD.90.115004
  [arXiv:1409.7690 [hep-ph]].
\bibitem{Dorsner:2015mja} 
  I.~Dor¨ner, S.~Fajfer, A.~Greljo, J.~F.~Kamenik, N.~Ko¨nik and I.~Ni¨and¸ic,
  JHEP {\bf 1506}, 108 (2015)
  doi:10.1007/JHEP06(2015)108
  [arXiv:1502.07784 [hep-ph]].
\bibitem{Crivellin:2015hha} 
  A.~Crivellin, J.~Heeck and P.~Stoffer,
  arXiv:1507.07567 [hep-ph].
\bibitem{He:2015rqa} 
  X.~G.~He, J.~Tandean and Y.~J.~Zheng,
  JHEP {\bf 1509}, 093 (2015)
  doi:10.1007/JHEP09(2015)093
  [arXiv:1507.02673 [hep-ph]].
\bibitem{Altmannshofer:2015esa} 
  W.~Altmannshofer, S.~Gori, A.~L.~Kagan, L.~Silvestrini and J.~Zupan,
  arXiv:1507.07927 [hep-ph].
\bibitem{Kosnik:2015lka} 
  N.~Ko¨nik,
  arXiv:1509.04590 [hep-ph].
\bibitem{Omura:2015xcg} 
  Y.~Omura, E.~Senaha and K.~Tobe,
  arXiv:1511.08880 [hep-ph].
\bibitem{Benbrik:2015evd} 
  R.~Benbrik, C.~H.~Chen and T.~Nomura,
  arXiv:1511.08544 [hep-ph].
  \bibitem{Bizot:2015qqo}
  N.~Bizot, S.~Davidson, M.~Frigerio and J.-L.~Kneur,
  arxiv:1512.08508.
\bibitem{Cheng:1987rs} 
  T.~P.~Cheng and M.~Sher,
  Phys.\ Rev.\ D {\bf 35}, 3484 (1987).
  doi:10.1103/PhysRevD.35.3484
\bibitem{Branco:1996bq} 
  G.~C.~Branco, W.~Grimus and L.~Lavoura,
  Phys.\ Lett.\ B {\bf 380}, 119 (1996)
  doi:10.1016/0370-2693(96)00494-7
  [hep-ph/9601383].
\bibitem{Botella:2015hoa} 
  F.~J.~Botella, G.~C.~Branco, M.~Nebot and M.~N.~Rebelo,
  arXiv:1508.05101 [hep-ph].
\bibitem{Altunkaynak:2015twa} 
  B.~Altunkaynak, W.~S.~Hou, C.~Kao, M.~Kohda and B.~McCoy,
  Phys.\ Lett.\ B {\bf 751}, 135 (2015)
  doi:10.1016/j.physletb.2015.10.024
  [arXiv:1506.00651 [hep-ph]].
    \bibitem{Buschmann:2016uzg}
     M. ~Buschmann, J. ~Kopp, J. ~Liu and X.-P. ~Wang, arxiv 1601.02616 [hep-ph]
\bibitem{Harnik:2012pb} 
  R.~Harnik, J.~Kopp and J.~Zupan,
  JHEP {\bf 1303}, 026 (2013)
  doi:10.1007/JHEP03(2013)026
  [arXiv:1209.1397 [hep-ph]].
\bibitem{Khachatryan:2014wca} 
  V.~Khachatryan {\it et al.} [CMS Collaboration],
  JHEP {\bf 1410}, 160 (2014)
  doi:10.1007/JHEP10(2014)160
  [arXiv:1408.3316 [hep-ex]].
\bibitem{Aad:2014vgg} 
  G.~Aad {\it et al.} [ATLAS Collaboration],
  JHEP {\bf 1411}, 056 (2014)
  doi:10.1007/JHEP11(2014)056
  [arXiv:1409.6064 [hep-ex]].
\cite{Paschos:1976ay}
\bibitem{Paschos:1976ay} 
  E.~A.~Paschos,
  Phys.\ Rev.\ D {\bf 15}, 1966 (1977).
  doi:10.1103/PhysRevD.15.1966
\bibitem{Glashow:1976nt} 
  S.~L.~Glashow and S.~Weinberg,
  Phys.\ Rev.\ D {\bf 15}, 1958 (1977).
  doi:10.1103/PhysRevD.15.1958
\bibitem{Davidson:2005cw} 
  S.~Davidson and H.~E.~Haber,
  Phys.\ Rev.\ D {\bf 72}, 035004 (2005)
  [Phys.\ Rev.\ D {\bf 72}, 099902 (2005)]
  doi:10.1103/PhysRevD.72.099902, 10.1103/PhysRevD.72.035004
  [hep-ph/0504050].
\bibitem{Mahmoudi:2009zx} 
  F.~Mahmoudi and O.~Stal,
  Phys.\ Rev.\ D {\bf 81}, 035016 (2010)
  doi:10.1103/PhysRevD.81.035016
  [arXiv:0907.1791 [hep-ph]].
\bibitem{Craig:2013hca} 
  N.~Craig, J.~Galloway and S.~Thomas,
  arXiv:1305.2424 [hep-ph].
\bibitem{Babu:2002et} 
  K.~S.~Babu and C.~Kolda,
  Phys.\ Rev.\ Lett.\  {\bf 89}, 241802 (2002)
  doi:10.1103/PhysRevLett.89.241802
  [hep-ph/0206310].
\bibitem{Sher:2002ew} 
  M.~Sher,
  Phys.\ Rev.\ D {\bf 66}, 057301 (2002)
  doi:10.1103/PhysRevD.66.057301
  [hep-ph/0207136].


  \end{thebibliography}
\end{document}